\documentclass[showpacs,prc,preprint,nofootinbib,showkeys]{revtex4}
\pdfoutput=1

\usepackage{graphicx}
\usepackage{amsmath}
\usepackage{amssymb}
\usepackage{bm}
\usepackage[usenames]{color}
\usepackage[colorlinks=false,linktocpage=true]{hyperref}

\newcommand{\beq}{\begin{equation}}
\newcommand{\eeq}{\end{equation}}
\newcommand{\bqa}{\begin{eqnarray}}
\newcommand{\eqa}{\end{eqnarray}}

\newcommand{\rs}{{\rm a}}

\newcommand{\J}{\tilde{{\cal J}}}

\newcommand{\I}{\tilde{{\cal I}}}

\newcommand{\pit}{\tilde{\pi}}
\newcommand{\Pit}{\tilde{\Pi}}
\newcommand{\V}{\tilde{V}}

\newcommand{\dft}{\delta\tilde{f}}

\begin{document}


\title{Nonconformal viscous anisotropic hydrodynamics}

\author{Dennis Bazow}
\affiliation{Department of Physics, The Ohio State University,
  Columbus, OH 43210 United States}
  
\author{Ulrich Heinz}
\affiliation{Department of Physics, The Ohio State University,
  Columbus, OH 43210 United States}

\author{Mauricio Martinez}
\affiliation{Department of Physics, The Ohio State University,
  Columbus, OH 43210 United States}

\begin{abstract}
We generalize the derivation of viscous anisotropic hydrodynamics from kinetic theory to allow for non-zero particle masses. The macroscopic theory is obtained by taking moments of the Boltzmann equation after expanding the distribution function around a spheroidally deformed local momentum distribution whose form has been generalized by the addition of a scalar field that accounts non-perturbatively (i.e. already at leading order) for bulk viscous effects. Hydrodynamic equations for the parameters of the leading-order distribution function and for the residual (next-to-leading order) dissipative flows are obtained from the three lowest moments of the Boltzmann equation. The approach is tested for a system undergoing (0+1)-dimensional boost-invariant expansion for which the exact solution of the Boltzmann equation in relaxation time approximation is known. Nonconformal viscous anisotropic hydrodynamics is shown to approximate this exact solution more accurately than any other known hydrodynamic approximation. 
\end{abstract}


\pacs{12.38.Mh, 25.75.-q, 24.10.Nz, 52.27.Ny, 51.10.+y} 
\keywords{Quark-Gluon Plasma, Anisotropic Dynamics, Boltzmann Equation, Viscous Hydrodynamics}
\date{\today }

\maketitle 


\section{Introduction}
\label{sec:introduction}
Relativistic fluid dynamics has been extensively used to describe the soft collective motion of relativistic heavy-ion collisions (see, for instance, \cite{Romatschke:2009im,Heinz:2013th,Gale:2013da} and references therein) which plays a central role in the phenomenology of the quark-gluon plasma. This led to a number of works aimed at exploring the theoretical foundations of relativistic fluid dynamics, with the goal of identifying improved hydrodynamic approximations of the underlying microcopic dynamics \cite{Muronga:2006zx,Betz:2010cx,Betz:2009zz,Denicol:2010xn,Denicol:2012cn,Denicol:2012es, Jaiswal:2013npa,Jaiswal:2013vta,Bazow:2013ifa}. For systems of deconfined quarks and gluons, the shear viscosity is expected to be much larger than the bulk viscosity at very high temperatures. It is then typically assumed that the bulk viscous pressure, whose Navier-Stokes value is porportional to the bulk viscosity, can be ignored in applications to heavy-ion physics. However, at temperature regimes typically produced experimentally in heavy-ion collisions, the order of magnitude of the bulk viscosity is unknown, and near the quark-hadron phase transition it could be large, due to the breaking of scale invariance by critical fluctuations and correlations \cite{Karsch:2007jc,Moore:2008ws}. Therefore, it is not well justified to a {\it priori} neglect the bulk viscous pressure when modeling the dynamics of QCD matter created at the Relativistic Heavy Ion Collider (RHIC) at Brookhaven National Laboratory and the Large Hadron Collider (LHC) at CERN. Indeed, several recent studies discussed the possibility of non-negligible bulk viscous effects on heavy-ion observables \cite{Torrieri:2008ip, Monnai:2009ad, Denicol:2009am, Song:2009rh, Rajagopal:2009yw, Bozek:2009dw,Bozek:2011ua,Dusling:2011fd,Bozek:2012qs, Noronha-Hostler:2013gga, Noronha-Hostler:2014dqa}. One reason for possibly larger than originally expected bulk viscous effects in heavy-ion collisions is the existence of shear-bulk couplings in the equations of motion that control the evolution of the shear and bulk vicous pressures \cite{Denicol:2010xn,Denicol:2012cn, Denicol:2012es}. Heavy-ion collisions are characterized by initially very large differences between the longitudinal and transverse expansion rates that cause large shear stress which, in turn, generates bulk viscous pressure via bulk-shear coupling \cite{Denicol:2014mca}. This mechanism should be included in phenomenological applications.

A key assumption made in hydrodynamics is that the system remains close to local equilibrium. This assumption breaks down during the very early expansion stage of the systems formed in ultrarelativistic heavy-ion collisions. To account for these large early-time deviations from local momentum isotropy, a framework called ``anisotropic hydrodynamics'' ({\sc aHydro}) was developed \cite{Martinez:2010sc, Florkowski:2010cf} and recently generalized to ``viscous (or second-order) anisotropic hydrodynamics'' ({\sc vaHydro}) \cite{Bazow:2013ifa}. In anisotropic hydrodynamics, one expands around an anisotropic background $f_\rs$ where the largest local momentum-space anisotropies are built already into the leading-order (LO) term:
\begin{equation}
   f(x,p) = 
   f_\rs\!\left(\frac{\sqrt{p^\mu \Xi_{\mu\nu}(x)p^\nu}}{\Lambda(x)} \right)
   + \dft(x,p).
\label{eq:faniso}
\end{equation}
Here $\Xi_{\mu\nu}$ is a second-rank tensor whose structure depends on the shape and amount of the  momentum-space anisotropy in the local fluid rest frame, and $\Lambda$ is a temperature-like scale which reduces to the true local temperature in the isotropic equilibrium limit. Ref.~\cite{Bazow:2013ifa} was the first to include the correction $\dft$ (which was computed using a Grad-Isreal-Stewart 14-moment approximation), leading to the equations of viscous anisotropic hydrodynamics ({\sc vaHydro}). When taking moments of the Boltzmann equation with the ansatz (\ref{eq:faniso}), the contributions from $\dft$ lead to additional dissipative (irreversible) currents $\Pit$ and $\pit^{\mu\nu}$ that account for local momentum anisotropies not already built into the leading-order distribution function $f_\rs(\sqrt{p{\cdot}\Xi{\cdot}p}/\Lambda)$ and their relaxation equations \cite{Bazow:2013ifa}. In \cite{Bazow:2013ifa} these equations were eventually simplified by assuming massless degrees of freedom which allowed the longitudinal and transverse components ${\cal P}_L$ and ${\cal P}_\perp$ of the anisotropic local pressure to factor into an isotropic thermal equilibrium pressure ${\cal P}_\mathrm{iso}$ multiplied by longitudinal and transverse ``deformation factors'' ${\cal R}_{L,\perp}(\xi)$ which depend only on the local momentum anisotropy parameter $\xi$. Of course, this assumption also implied zero bulk viscous pressure $\Pi$. As a test of the {\sc vaHydro} approach these simplified equations were then solved numerically for a transversally homogeneous system undergoing boost-invariant longitudinal expansion ((0{+}1)-dimensional expansion) for which the underlying Boltzmann equation can be solved exactly in the relaxation time approximation (RTA). Comparison of this exact solution with {\sc vaHydro} as well as several other viscous hydrodynamic approximations revealed a uniformly superior performance of the {\sc vaHydro} scheme.

We here generalize the {\sc vaHydro} approach to the massive case where the above simplifications no longer hold. A suitable generalization of the tensor $\Xi^{\mu\nu}(x)$ in Eq.~(\ref{eq:faniso}) was first written down in \cite{Martinez:2012tu} but not immediately exploited. It was recently shown that, with this generalization, already LO anisotropic hydrodynamics (without the $\dft$ terms) implicitly contains {\it some} of the shear-bulk couplings present in modern versions of second-order viscous hydrodynamics \cite{Nopoush:2014pfa,Denicol:2014mca}. By generalizing the work \cite{Bazow:2013ifa} to the massive case, we here extend the works \cite{Nopoush:2014pfa,Denicol:2014mca} to next-to-leading order ({\it i.e.} we generalize the non-conformal {\sc aHydro} approximation used in \cite{Nopoush:2014pfa, Denicol:2014mca} to non-conformal {\sc vaHydro}), keeping all additional shear and bulk viscous corrections arising from the $\dft$ term in Eq.~(\ref{eq:faniso}). This improved approach is again tested in a system undergoing (0+1)-dimensional boost-invariant expansion for which the exact solution of the RTA Boltzmann equation was recently extended to a gas of massive particles \cite{Florkowski:2014sfa}.

The structure of this paper is as follows. In Sec.~\ref{sec:kinetic_theory} we briefly review how to derive the macroscopic hydrodynamic variables from kinetic theory by expanding the local distribution function around a spheroidal momentum distribution. In Sec.~\ref{sec:distrexp} we write down the form of the leading order anisotropic background as well as the perturbations around this form. Then in Sec.~\ref{sec:equations_of_motion} we present the equations of motion for non-conformal viscous anisotropic hydrodynamics. In Sec.~\ref{sec:residual_moments} we summarize the basic steps applied to obtain a closed form expression of the dynamical equations of motion for the residual dissipative currents, using the macroscopic conservation laws. Sec.~\ref{sec:0+1d.} we simplify these equations for the limiting case of a (0+1)-dimensional longitudinally boost-invariant expansion with finite particle masses and compare their numerical solution to the exact result from solving the microscopic Boltzmann equation. In Sec.~\ref{sec:concl.} we present our conclusions.

Our notation is as follows: We use natural units $\hbar=k_B=c=1$. The Minkowski metric tensor is $g^{\mu\nu}={\rm diag}(+,-,-,-)$. Greek indices run from 0 to 3 and Latin indices from 1 to 3. The summation convention for repeated indices (Greek or Latin) is always used. Our tensor basis, in the local rest frame, is $X^\mu_0\equiv u^\mu=(1,0,0,0)$, $X^\mu_1\equiv x^\mu=(0,1,0,0)$, $X^\mu_2\equiv y^\mu=(0,0,1,0)$, and $X^\mu_3\equiv z^\mu=(0,0,0,1)$. The transverse projection operator $\Delta^{\mu\nu}\equiv {-}X^\mu_iX^\nu_i=g^{\mu\nu}{-}u^\mu u^\nu$ is used to project four-vectors and/or tensors into the space orthogonal to $u^\mu$. The notations $A^{(\mu\nu)}\equiv\frac{1}{2}\left(A^{\mu\nu}{+}A^{\nu\mu}\right)$ and $A^{[\mu\nu]}\equiv\frac{1}{2}\left(A^{\mu\nu}{-}A^{\nu\mu}\right)$ denote symmetrization and antisymmetrization, respectively. $A^{\langle \mu \nu\rangle}\equiv\Delta^{\mu\nu}_{\alpha\beta}A^{\alpha\beta}$ where $\Delta^{\mu\nu}_{\alpha\beta}\equiv\Delta^{(\mu}_\alpha\Delta^{\nu)}_\beta-\Delta^{\mu\nu}\Delta_{\alpha\beta}/3$ is the transverse (to $u$) and traceless projector for second-rank tensors. The four-derivative is $\partial_\mu\equiv\partial/\partial x^\mu$, $D\equiv u^\mu\partial_\mu$ is the convective derivative (the time derivative in the comoving frame), $\nabla^\mu\equiv\Delta^{\mu\nu}\partial_\nu$ is the covariant notation for the spatial gradient operator in the local rest frame, and $\theta\equiv\partial_\mu u^\mu=\nabla_\mu u^\mu$ is the scalar expansion rate.

\section{Hydrodynamics from relativistic kinetic theory}
\label{sec:kinetic_theory}

To keep the presentation selfcontained, we here briefly review how to extract hydrodynamic variables from the Boltzmann equation for an expansion around a locally spheroidal momentum distribution \cite{Bazow:2013ifa}. In kinetic theory the one-particle distribution function $f$ is governed by the Boltzmann equation,
\begin{equation}
p^\mu\partial_\mu f=C[f] \label{eq:be} \; ,
\end{equation}
where $C[f]$ is the collision kernel. The particle current and energy-momentum tensor are expressed as the first and second moments of the one-particle distribution function
\begin{equation}
J^\mu=\langle p^\mu\rangle\;,\;\;\;\;\; T^{\mu\nu}=\langle p^\mu p^\nu\rangle\;,
\end{equation}
where we defined the average of a momentum-dependent observable ${\cal O}(p)$ at point $x$ as
\begin{equation}
\langle{\cal O}\rangle(x)\equiv\int dP\,{\cal O}(p)f(x,p)
\end{equation}
with the Lorentz invariant momentum-space measure $dP\equiv (2\pi)^{-3}(d^3p/E)$. We decompose the particle four-momentum $p^\mu$ into parts parallel and orthogonal to the four-velocity $u^\mu$ of the local fluid rest frame \cite{Anderson:1974},
\begin{equation} 
p^\mu= E u^\mu+p_i X^\mu_i , \label{eq:p_dec_aniso}
\end{equation}
where $E{\,=\,}p{\cdot}u(x)$ is the local rest frame energy and $p_i{\,=\,}- X_i^\nu(x)p_\nu$ are the Cartesian components of the four-momentum in the local rest frame. For systems that are locally approximately spheroidal in momentum-space, characterized by a ``local anisotropic equilibrium'' distribution function $f_\rs$, we can decompose $f$ as
\begin{equation}
f(x,p)=f_\rs(x,p)+\dft\,.
\end{equation}
Then $J^\mu$ and $T^{\mu\nu}$ can be tensor decomposed as
\begin{align}
\label{JmuTmunu}
J^\mu&={\cal N}u^\mu+\V^\mu\;,\\
T^{\mu\nu}&={\cal E}u^\mu u^\nu-({\cal P}_\perp+\Pit)\Delta^{\mu\nu}
+\left({\cal P}_L-{\cal P}_\perp\right)z^\mu z^\nu+\pit^{\mu\nu}\;.
\end{align}
Here $z^\mu(x){\,\equiv\,}X^\mu_3(x)$ is the four-vector that reduces in the local fluid rest frame to a unit vector in longitudinal ($z$) direction, ${\cal N}$ is the particle density and $\V^\mu$ is the particle current in the local rest frame, ${\cal E}$ is the energy density in the local rest frame, ${\cal P}_\perp$ and ${\cal P}_L$ are the transverse and longitudinal pressures, $\Pit$ is the bulk viscous pressure, and $\pit^{\mu\nu}$ is the shear stress tensor defined by
\begin{equation} 
\begin{array}{lll}
\label{dissipativeCurrents}
{\cal N}\equiv\langle E\rangle_\rs, & \V^\mu\equiv \langle p_i\rangle_{\tilde\delta}X^\mu_i, & {}\\
{\cal E}\equiv \langle E^2 \rangle_\rs, & {\cal P}_\perp\equiv \langle p^2_\perp \rangle_\rs & {\cal P}_L\equiv \langle p^2_z \rangle_\rs 
\\
\Pit\equiv -\frac{1}{3}\langle \Delta^{\alpha\beta}p_\alpha\ p_\beta\rangle_{\tilde\delta},  & \pi^{\mu\nu}\equiv\langle p^{\langle \mu}p^{\nu \rangle}\rangle_{\tilde\delta} & {}\;.
\end{array}  
\end{equation}
In these equations we introduced the notation $\langle\cdots\rangle_\rs\equiv\int dP\,(\cdots)f_\rs$ and $\langle\cdots\rangle_{\tilde\delta}\equiv\int dP\,(\cdots)\dft$, and made use of the generalized Landau matching conditions $\langle E\rangle_{\tilde\delta}=\langle E^2\rangle_{\tilde\delta}=0$. For later convenience, the total bulk viscous pressure $\Pi$ is calculable as
\begin{equation}
\label{eq28}
\Pi = \frac{2{\cal P}_\perp+{\cal P}_L}{3}-{\cal P}_\mathrm{eq}+\Pit
    \;,
\end{equation}
and the total shear stress tensor is obtained from
\begin{eqnarray}
\label{eq29a}
\pi^{\mu\nu} &=&  \left({\cal P}_L{-}{\cal P}_\perp\right)
                            \left(\frac{\Delta^{\mu\nu}}{3}+z^\mu z^\nu\right) + \pit^{\mu\nu} 
= \left({\cal P}_\perp{-}{\cal P}_L\right)\,\frac{x^\mu x^\nu{+}y^\mu y^\nu{-}2 z^\mu z^\nu}{3}
       + \pit^{\mu\nu} . 
\end{eqnarray}

\section{14-moment approximation for the distribution function}
\label{sec:distrexp}

\subsection{Leading order (LO) distribution function}
\label{sec:lodist}

In this paper we consider systems that are, to leading order, spheroidal in momentum-space in the local rest frame. This is accomplished by introducing the anisotropy tensor $\Xi^{\mu\nu}$, so that the leading-order one-particle distribution function takes the form:
\begin{equation}
f_\rs=f_0\left(\frac{1}{\Lambda(x)}\sqrt{p^\mu\Xi_{\mu\nu}(x)p_\nu}\right) , 
\label{eq:lo_df}
\end{equation}
where we have assumed zero chemical potential and $f_0$ has the functional form of a local thermal equilibrium distribution,
\begin{equation}
f_0(y)\equiv\frac{1}{e^{y}+a}\;,
\end{equation}
where $a=\pm1,0$ corresponds to Fermi-Dirac, Bose-Einstein, and classical Boltzmann statistics, respectively.

The most general decomposition of the rank-two tensor $\Xi^{\mu\nu}$ that possesses spheroidal symmetry in the local rest frame is~\cite{Martinez:2012tu}
\begin{equation}
\Xi^{\mu\nu}=u^\mu u^\nu -\Phi\Delta^{\mu\nu}+\xi z^{\mu}z^{\nu}\;,
\end{equation}
where all terms are functions of position $x$. In local rest frame coordinates $f_\rs$ takes the form
\begin{equation}
f_\rs=f_0\left(\frac{1}{\Lambda}\sqrt{m^2{+}(1{+}\Phi)p^2_\perp{+}(1{+}\Phi{+}\xi)p^2_z}\right) 
\equiv f_0\left(\frac{E_\rs}{\Lambda}\right), 
\label{eq:lo_df2}
\end{equation}
where we defined $E_\rs^2{\,\equiv\,}(1{+}\Phi)m_\perp^2\cosh^2y + \xi m_\perp^2\sinh^2y - \Phi m^2$, with $m_\perp^2{\,=\,}m^2{+}p_\perp^2$. Connecting the ``anisotropic equilibrium" quantities with moments of $f_\rs$, one finds that these quantities can be written as (see Appendix)
\begin{align}
&{\cal N}\left(\Lambda,\xi,\Phi;\hat{m}\right)=\langle E\rangle_\rs=\frac{{\cal N}_0(\Lambda)}{(1{+}\Phi)(1{+}\Phi{+}\xi)}\,,
\label{anisoN}\\
&{\cal E}\left(\Lambda,\xi,\Phi;\hat{m}\right)=\langle E^2\rangle_\rs 
\nonumber\\
&= 
\frac{\Lambda^4}{2\pi^2}\int_0^\infty\!\! dy\,\cosh^{2}y
\int_{\hat{m}}^\infty d\hat{m}_\perp\,\hat{m}_\perp^3\, 
       f_0\Bigl(\sqrt{\hat{m}_\perp^2[(1{+}\Phi)\cosh^2y+\xi\sinh^{2}y]-\hat{m}^2\Phi}\Bigr),
\label{anisoE}\\
&{\cal P}_\perp\left(\Lambda,\xi,\Phi;\hat{m}\right)=\langle p_\perp^2\rangle_\rs 
\nonumber\\
&=
\frac{\Lambda^4}{4\pi^2}\int_0^\infty \!\! dy
\int_{\hat{m}}^\infty d\hat{m}_\perp\,\hat{m}_\perp
\left(\hat{m}_\perp^2{-}\hat{m}^2\right)
      f_0\Bigl(\sqrt{\hat{m}_\perp^2[(1{+}\Phi)\cosh^2y+\xi\sinh^{2}y]-\hat{m}^2\Phi}\Bigr),
\\
&{\cal P}_L\left(\Lambda,\xi,\Phi;\hat{m}\right)=\langle p^2_z\rangle_\rs 
\nonumber\\
&=
\frac{\Lambda^4}{2\pi^2}\int_0^\infty\!\! dy\,\sinh^{2}y
\int_{\hat{m}}^\infty d\hat{m}_\perp\,\hat{m}_\perp^3\,
   f_0\Bigl(\sqrt{\hat{m}_\perp^2[(1{+}\Phi)\cosh^2y+\xi\sinh^{2}y]-\hat{m}^2\Phi}\Bigr),
\end{align}
where $\hat{m}{\,\equiv\,}m/\Lambda$ and $\hat{m}_\perp{\,\equiv\,}m_\perp/\Lambda{\,\equiv\,}\sqrt{m^2{+}p_\perp^2}/\Lambda$. The equilibrium thermodynamic quantities are given as moments of $f_{0}$. For Boltzmann statistics they have the functional form:
\begin{align}
{\cal N}_0(T;m)&\equiv\frac{Tm}{2\pi^2}K_{2}(m/T)\;,\label{isoN}\\
{\cal E}_0(T;m)&\equiv\frac{T^2m^2}{2\pi^2}\left(3K_{2}(m/T)+\frac{m}{T}K_{1}(m/T)\right)\;,\\
{\cal P}_0(T;m)&\equiv{\cal N}(T;m)T\;,
\end{align}
where $K_n(z)$ are the modified Bessel function of the second kind.
%

\subsection{14-moment expansion of the deviation from the LO distribution}
\label{sec:deltaf}

In the 14-moment approximation, the deviation $\dft$ of the full distribution function $f$ from the locally anisotropic state (\ref{eq:lo_df}) is expanded to second order in momenta as \cite{Bazow:2013ifa}:
\begin{equation} \label{q_expanded}
\begin{split}
\frac{\dft}{f_\rs\tilde{f}_\rs} &= \alpha - \beta E + w E^2 - \frac{w}{3} \Delta^{\mu \nu} p_{\mu} p_{\nu}
+ w_{\langle \mu \nu \rangle} p^{\langle\mu} p^{\nu\rangle}\, ,
\end{split}
\end{equation}
where $\tilde{f}_\rs{\,\equiv\,}1{-}af_\rs$. In the absence of a chemical potential, as assumed in (\ref{eq:lo_df}), there is no heat current $\tilde{V}^\mu$, and the coefficients of any terms linear in $p^{\langle\mu\rangle}$ in Eq.~(\ref{q_expanded}) vanish. By inserting Eq.~(\ref{q_expanded}) into the definitions (\ref{dissipativeCurrents}) of the residual dissipative flows, the 14-moment coefficients can be expressed in terms of these flows by solving the matrix equation ${\cal A} \bm{b}=\bm{c}$, where
\begin{eqnarray}
\label{79}
&&{\cal A}\equiv 
 \begin{pmatrix}
\J_{1,0} & -\J_{2,0} & \J_{3,0}+\J_{3,1} &0&0&0&0& 0 &\rho^{zz}_{1,0} 
\\
\J_{2,0} & -\J_{3,0} & \J_{4,0}+\J_{4,1} &0&0&0&0& 0 &\rho^{zz}_{2,0}
\\
\J_{2,1} & -\J_{3,1} & \J_{4,1}+\frac{5}{3}\J_{4,2} &0&0&0&0& 0 &\rho^{zz}_{2,1}
\\  
\varphi^{xx}_{21} & -\varphi^{xx}_{31} & \varphi^{xx}_{41}+\varphi^{xx}_{42}&  \lambda^{1111} & 0 & 0 & \lambda^{1122} & 0 & \lambda^{1133} 
\\
0&0&0&0 & 2\lambda^{1212} & 0 & 0 & 0 & 0 
\\
0&0&0& 0  & 0  & 2\lambda^{1313} & 0 & 0 & 0  
\\
\varphi^{xx}_{21} & -\varphi^{xx}_{31} & \varphi^{xx}_{41}+\varphi^{xx}_{42}&  \lambda^{1122} & 0  & 0 & \lambda^{1111} & 0 & \lambda^{1133} 
\\
0&0&0&  0  & 0  & 0  & 0  & 2\lambda^{1313} & 0 
\\
\varphi^{zz}_{21} & -\varphi^{zz}_{31} & \varphi^{zz}_{41}+\varphi^{zz}_{42}&  \lambda^{1133} & 0  & 0 & \lambda^{1133} & 0 & \lambda^{3333}
 \end{pmatrix} ,
\\ \label{80}
&&\bm{b}\equiv 
 \begin{pmatrix}
  \alpha & \beta & w & w_{11} & w_{12} & w_{13} & w_{22} & w_{23} & w_{33} 
 \end{pmatrix}^T \,,
\\ \label{81}
&&\bm{c}\equiv 
 \begin{pmatrix}
  0& 0 & \Pit & \pit_{11} & \pit_{12} & \pit_{13} & \pit_{22} & \pit_{23} & \pit_{33} 
 \end{pmatrix}^T \,.
\end{eqnarray}
This allows the distribution function expanded around an anisotropic background to be expressed in terms of the residual dissipative flows $\Pit$ and $\pit^{\mu\nu}$ as
\begin{equation}
\label{eq:f14moment}
f=f_\rs+\left[\lambda_\Pi\Pit+\lambda^{\mu\nu}_\pi\pit_{\mu\nu}
+\left(\lambda_\Pi^{\mu\nu}\Pit
+\lambda_\pi^{\mu\nu\alpha\beta}\pit_{\alpha\beta}\right)
p_{\langle\mu}p_{\nu\rangle}\right]f_\rs\tilde{f}_\rs\;,
\end{equation}
where $\lambda_\Pi$, $\lambda^{\mu\nu}_\pi$, $\lambda^{\mu\nu}_\Pi$, and $\lambda^{\alpha\beta\mu\nu}_\pi$, along with the auxiliary tensors $\rho^{\mu\nu}_{nq}$, $\varphi^{\alpha\beta}_{nq}$, and $\lambda^{\mu\nu\alpha\beta}$ appearing in (\ref{79}), are defined in Ref.~\cite{Bazow:2013ifa}.

\section{Viscous anisotropic hydrodynamic equations of motion}
\label{sec:equations_of_motion}
In this section we derive the hydrodynamic equations of motion by taking moments of the Boltzmann equation. Taking moments implies multiplying (\ref{eq:be}) by integer powers of the four-momentum and integrating over momentum-space. This process results in the following $n$-th ($n\ge 0$) moment equation:
\begin{equation}
\label{infMom}
\partial_{\mu_1} \langle p^{\mu_1}\cdots p^{\mu_{n+1}} \rangle={\cal C}^{\mu_1\cdots \mu_{n}}\;. 
\end{equation}
The $n$-th rank collisional tensor is defined in the following manner:
\begin{equation}
{\cal C}^{\mu_1 \cdots \mu_n}_r=
\int dP\,E^rp^{\mu_1}\cdots p^{\mu_n} C[f]\,,
\end{equation}
with ${\cal C}^{\mu_1 \cdots \mu_n}\equiv{\cal C}_0^{\mu_1 \cdots \mu_n}$.  The infinite set of coupled moments (\ref{infMom}) is equivalent to knowing the full solution $f$ of the Boltzmann equation (\ref{eq:be}). Only the first few moments have an intuitive physical meaning: The zeroth moment $\partial_\mu\langle p^\mu\rangle={\cal C}$ embodies the conservation of particle number for vanishing ${\cal C}$, the first moment $\partial_\mu\langle p^\mu p^\nu\rangle=0$ the conservation of energy and momentum. The macroscopic equations of spheroidal viscous anisotropic hydrodynamics are derived in the following subsections.

\subsection{Zeroth moment of the Boltzmann equation}
\label{subsec:bem0}
The zeroth moment of the Boltzmann equation gives
\begin{equation}
\partial_\mu j^\mu = D{\cal N}+{\cal N}\theta+\partial_\mu \V^\mu={\cal C} \label{eq:be0_tmp} .
\end{equation}
Denoting the action of the time derivative in the local rest frame $D$ by a dot, Eq.~(\ref{eq:be0_tmp}) can be written as an equation of motion for the rest frame particle density ${\cal N}$:
\begin{equation}
\label{eq:be0}
\dot{\cal N}=-{\cal N}\theta-\partial_\mu \V^\mu+{\cal C}.
\end{equation}
%
\subsection{First moment of the Boltzmann equation}
\label{subsec:bem1}
The first moment of the Boltzmann equation is equivalent to the requirement of energy-momentum conservation: $\partial_\mu T^{\mu\nu}=0$. With the viscous anisotropic hydrodynamic decomposition of $T^{\mu\nu}$ given in (\ref{JmuTmunu}) this conservation law yields
\begin{equation}
\begin{split}
\partial_\mu T^{\mu\nu}&=u^\nu D({\cal E}{+}{\cal P}_\perp{+}\Pit)
  + u^\nu ({\cal E}{+}{\cal P}_\perp{+}\Pit)\theta
  + ({\cal E}{+}{\cal P}_\perp{+}\Pit) D u^\nu
   - \partial^\nu({\cal P}_\perp{+}\Pit) \\
&+ z^\nu D_L({\cal P}_{\rm L}{-}{\cal P}_\perp)
  + z^\nu({\cal P}_{\rm L}{-}{\cal P}_\perp)\theta_{\rm L} 
  + ({\cal P}_{\rm L}{-}{\cal P}_\perp)D_L z^\nu + \partial_\mu\pit^{\mu\nu}=0.
\end{split}
\end{equation}
Projecting these four equations on the fluid four-velocity yields an equation of motion for the rest frame energy density ${\cal E}$: 
\begin{equation}
u_\nu\partial_\mu T^{\mu\nu}=\dot{{\cal E}} + ({\cal E}{+}{\cal P}_\perp{+}\Pit)\theta
+ ({\cal P}_{\rm L}{-}{\cal P}_\perp)u_\nu D_z z^\nu-\pit^{\mu\nu}\sigma_{\mu\nu}=0.
\label{eq:par}
\end{equation}
The projections $\Delta^\alpha_{\ \nu} \partial_\mu T^{\mu\nu}$ transverse to $u^\mu$ yield equations of motion for the fluid four-velocity $u^\mu$:
\begin{eqnarray}
\label{eq:perp} 
&&\Delta^\alpha_{\ \nu} \partial_\mu T^{\mu\nu}
= ({\cal E}{+}{\cal P}_\perp{+}\Pit)\dot{u}^\alpha 
 - \nabla^\alpha ({\cal P}_\perp{+}\Pit)
+ \Delta^\alpha_{\ \nu}\partial_\mu\pit^{\mu\nu}
\\\nonumber
&&\qquad\qquad + z^\alpha D_z ({\cal P}_{\rm L}{-}{\cal P}_\perp)
+ z^\alpha ({\cal P}_{\rm L}{-}{\cal P}_\perp)(\partial_{\mu}z^{\mu}) 
+  ({\cal P}_{\rm L}{-}{\cal P}_\perp)D_z z^\alpha 
-({\cal P}_{\rm L}{-}{\cal P}_\perp)u^\alpha u_\nu D_z z^\nu=0.
\end{eqnarray}
In the above equations we have introduced the velocity shear tensor $\sigma^{\mu\nu}\equiv\nabla^{\langle\mu}u^{\nu\rangle}$ and the derivative operator $D_i\equiv X^\mu_i\partial_\mu$. Equations (\ref{eq:par}) and (\ref{eq:perp}) are the fundamental equations of relativistic viscous anisotropic hydrodynamics.
\subsection{Second moment of the Boltzmann equation}
\label{subsec:bem2}
The second moment of the Boltzmann equation gives
\begin{equation}
\label{2ndMom}
\partial_\mu{\cal F}^{\mu\nu\lambda}={\cal C}^{\nu\lambda}\;,
\end{equation}
where ${\cal F}^{\mu\nu\lambda}\equiv\langle p^\mu p^\nu p^\lambda\rangle$. Decomposing $p^\mu$ into parts parallel and orthogonal to $u^\mu$ by using Eq.~(\ref{eq:p_dec_aniso}) leads to 
\begin{eqnarray}
{\cal F}^{\mu\nu\lambda} &\equiv&
\langle E^3\rangle u^\mu u^\nu u^\lambda
+\langle E^2 p_i\rangle\left(u^\mu u^\nu X^\lambda_i+u^\mu X^\nu_i u^\lambda +X^\mu_i u^\nu u^\lambda\right)\nonumber\\
&+&\langle E p_ip_j\rangle\left(u^\mu X^\nu_i X^\lambda_j+X^\mu_i u^\nu X^\lambda_j +X^\mu_i X^\nu_j u^\lambda\right)
+\langle p_ip_jp_k\rangle X^\mu_i X^\nu_j X^\lambda_k\;.
\end{eqnarray}
To evaluate the l.h.s. of Eq.~(\ref{2ndMom}) requires taking the four-divergence of tensor ${\cal F}^{\mu\nu\lambda}$
\begin{eqnarray}
\partial_\mu{\cal F}^{\mu\nu\lambda} &=&
D\langle E^3\rangle u^\nu u^\lambda
+\langle E^3\rangle\left(u^\nu u^\lambda\theta+2u^{(\nu}Du^{\lambda)}\right)\nonumber\\
&+&X^\nu_i X^\lambda_j D\langle Ep_ip_j\rangle
+\langle Ep_ip_j\rangle\left(X^\nu_i X^\lambda_j\theta+2X^{(\nu}_iDX^{\lambda)}_j\right)\nonumber\\
&+&2u^{(\nu}X^{\lambda)} D_i\langle Ep_ip_j\rangle
+2\langle Ep_ip_j\rangle\left(u^{(\nu}X^{\lambda)}_j\partial_\mu X^\mu_i+u^{(\nu}D_iX^{\lambda)}_j\right).
\end{eqnarray}
Projecting out the transverse to $u^\mu$ and traceless part of Eq.~(\ref{2ndMom}), $\Delta^{\alpha\beta}_{\nu\lambda}\partial_\mu{\cal F}^{\mu\nu\lambda}={\cal C}^{\langle\alpha\beta\rangle}$, yields
\begin{align}
\label{2ndMoment}
X^{\langle\alpha}_i X^{\beta\rangle}_j \left(D\langle E p_i p_j\rangle
+\langle E p_i p_j\rangle\theta\right)
+2\langle E p_i p_j\rangle\Delta^{\alpha\beta}_{\nu\lambda}
\left(X^{\nu}_i DX^{\lambda}_j
+X^{\nu}_i D_j u^{\lambda}\right)
={\cal C}^{\langle\alpha\beta\rangle}\;.
\end{align}
To work out the averages $\langle E p_i p_j\rangle$ appearing in Eq.~(\ref{2ndMoment}), we use (\ref{eq:f14moment}) to write
\begin{equation}
\label{epipj}
\langle E p_i p_j\rangle=\I^{ij}_{10}+\psi_\Pi^{ij}\Pit+\psi_\pi^{ij\mu\nu}\pit_{\mu\nu}\;,
\end{equation}
where $\I^{ij}_{10}$ is defined in the Appendix and 
\begin{align}
\psi_\Pi^{ij}&\equiv
\left[\int dP\,\lambda_{\Pi}Ep_{i}p_{j}f_\rs\tilde{f}_\rs
+\lambda^{\mu\nu}_{\Pi}\int dP\,p_{\langle\mu}p_{\nu\rangle}Ep_{i}p_{j}f_\rs\tilde{f}_\rs
\right]\Pit\;,
\\
\psi_\pi^{ij\mu\nu}&\equiv
\left[\int dP\,\lambda^{\mu\nu}_{\Pi}Ep_{i}p_{j}f_\rs\tilde{f}_\rs
+\lambda^{\alpha\beta\mu\nu}_{\Pi}\int dP\,p_{\langle\alpha}p_{\beta\rangle}Ep_{i}p_{j}f_\rs\tilde{f}_\rs
\right]\pit_{\mu\nu}\;.
\end{align}
Using (\ref{epipj}), equation~(\ref{2ndMoment}) can finally be written in the following form:
\begin{align}
\label{2ndMomentFinal}
&X^{\langle\alpha}_i X^{\beta\rangle}_j \left[\dot{\I}^{ij}_{10}
+\psi^{ij}_\Pi\dot{\Pit}+\psi^{ij\mu\nu}_\pi\dot{\pit}_{\mu\nu}
+\dot{\psi}^{ij}_\Pi\Pit+\dot{\psi}^{ij\mu\nu}_\pi\pit_{\mu\nu}
\right]
+X^{\langle\alpha}_i X^{\beta\rangle}_j \left[
\dot{\I}^{ij}_{10}
+\psi^{ij}_\Pi\Pit+\psi^{ij\mu\nu}_\pi\pit_{\mu\nu}
\right]\theta
\notag
\\&\hspace{2cm}
+2\left[
\dot{\I}^{ij}_{10}
+\psi^{ij}_\Pi\Pit+\psi^{ij\mu\nu}_\pi\pit_{\mu\nu}
\right]\Delta^{\alpha\beta}_{\nu\lambda}
\left(X^{\nu}_i DX^{\lambda}_j
+X^{\nu}_i D_j u^{\lambda}\right)
={\cal C}^{\langle\alpha\beta\rangle}\;.
\end{align}
%

\section{Equations of motion for the residual dissipative flows}
\label{sec:residual_moments}

To close the system of equations we need to know the space-time evolution of the dissipative currents appearing in (\ref{dissipativeCurrents}). To accomplish this, we derive the equations of motion for $\Pit$ and $\pit^{\mu\nu}$ from their kinetic definitions \cite{Denicol:2012cn,Denicol:2010xn,Bazow:2013ifa}:
\begin{eqnarray}
\dot{\Pit}&=&-\frac{m^2}{3}\int dP\,\dot{\dft}\;,\label{eq:evolve_res_moments}\\
\dot{\pit}^{\langle\mu\nu\rangle}&=&\Delta^{\mu\nu}_{\alpha\beta}\int dP\, p^{\langle\alpha}p^{\beta\rangle}\dot{\dft}\;.\label{eq:evolve_res_moments2}
\end{eqnarray}
The Boltzmann equation~(\ref{eq:be}) can be written in the form
\begin{equation} 
\label{eq:evolve_be}
\delta\dot{\tilde{f}}=-\dot{f}_\rs - \frac{1}{E}\Bigl(p{\cdot}\nabla (f_\rs{+}\dft)-C[f]\Bigr) .
\end{equation}
Substituting this into the expressions~(\ref{eq:evolve_res_moments}) and (\ref{eq:evolve_res_moments2}) one obtains the following equations of motion:
\begin{align}
\label{Pi1_0}
-\frac{3}{m^2}\dot\Pit-{\cal C}_{-1} &= {\cal W}-{\cal X}\theta-{\cal Y}^{\mu\nu}\sigma_{\mu\nu}+\frac{3}{m^2}\Pit\theta
-\langle E^{-2}p^\mu p^\nu\rangle_{\tilde\delta}\nabla_\mu u_\nu
\end{align}
\begin{align}
\label{pimunu1_0}
\dot{\pit}^{\left\langle \mu \nu \right\rangle }
-{\cal C}_{-1}^{\left\langle\mu \nu \right\rangle }
&= {\cal K}^{\mu\nu}+{\cal L}^{\mu\nu}+{\cal M}^{\mu\nu}
+{\cal H}^{\mu\nu\lambda}\left(\dot{z}_\lambda+u^\alpha\nabla_\lambda z_\alpha\right)
+(1+\Phi){\cal Q}^{\mu\nu\lambda\alpha}\nabla_\lambda u_\alpha \nonumber\\
&
-\frac{5}{3}\pit^{\mu\nu}\theta
-2\pit^{\langle\mu}_\lambda\sigma^{\nu\rangle\lambda} 
+2\pit^{\langle\mu}_\lambda\omega^{\nu\rangle\lambda}
+2\Pit\sigma^{\mu\nu} \nonumber\\
&
-\left\langle E^{-2}p^{\langle\mu}p^{\nu\rangle}p^{\langle\alpha\rangle}p^{\langle\beta\rangle}\right\rangle_{\tilde\delta}\nabla_\alpha u_\beta 
\;.
\end{align}
The evolution equations for the dissipative flows $\Pit$ and $\pit^{\mu\nu}$ can now be obtained by inserting the closed form (\ref{eq:f14moment}) of the single-particle distribution function into the expectation values $\langle \,\cdots\, \rangle_{\tilde\delta}$ on the r.h.s. of the equations of motion (\ref{Pi1_0}) and (\ref{pimunu1_0}). This leads to
\begin{align}
\label{Pi1}
-\frac{3}{m^2}\dot\Pit-{\cal C}_{-1} &= {\cal W}-{\cal X}\theta-{\cal Y}^{\mu\nu}\sigma_{\mu\nu}+\frac{3}{m^2}\Pit\theta
-\delta_{\Pi\Pi}^{\mu\nu}\Pit\nabla_\mu u_\nu-\pit_{\alpha\beta}\delta_{\Pi\pi}^{\mu\nu\alpha\beta}\nabla_\mu u_\nu
\end{align}
\begin{align}
\label{pimunu1}
\dot{\pit}^{\left\langle \mu \nu \right\rangle }
-{\cal C}_{-1}^{\left\langle\mu \nu \right\rangle }
&= {\cal K}^{\mu\nu}+{\cal L}^{\mu\nu}+{\cal M}^{\mu\nu}
+{\cal H}^{\mu\nu\lambda}\left(\dot{z}_\lambda+u^\alpha\nabla_\lambda z_\alpha\right)
+(1+\Phi){\cal Q}^{\mu\nu\lambda\alpha}\nabla_\lambda u_\alpha \nonumber\\
&
-\frac{5}{3}\pit^{\mu\nu}\theta
-2\pit^{\langle\mu}_\lambda\sigma^{\nu\rangle\lambda} 
+2\pit^{\langle\mu}_\lambda\omega^{\nu\rangle\lambda}
+2\Pit\sigma^{\mu\nu} \nonumber\\
&
-\Pit\delta_{\pi\Pi}^{\mu\nu\alpha\beta}\nabla_\alpha u_\beta
-\delta_{\pi\pi}^{\mu\nu\alpha\beta\sigma\lambda}\pit_{\sigma\lambda}\nabla_\alpha u_\beta
\;.
\end{align}
The dissipative forces for the bulk viscous pressure and shear-stress tensor evolution equation are defined as
\begin{align}
{\cal W}&\equiv\dot{\beta_\rs}\J_{0,0,1}+\frac{\beta_\rs}{2}\J^{zz}_{0,0,-1}\dot{\xi}+\frac{3}{2}\beta_\rs\J_{2,1,-1}\dot{\Phi}\;,\\
{\cal X}&\equiv\frac{\beta_\rs}{3}\left[
(1{+}\Phi)(2\J^{xx}_{0,0,-1}+\J^{zz}_{0,0,-1})+\xi\J^{zz}_{0,0,-1}
\right]\;,\\
{\cal Y}^{\mu\nu}&\equiv\left[
(1{+}\Phi)(\J^{zz}_{0,0,-1}-\J^{xx}_{0,0,-1})+\xi\J^{zz}_{0,0,-1}
\right]z^{\mu}z^{\nu}\;,\\
{\cal M}^{\mu\mu}&\equiv\frac{3\beta_\rs}{2}\dot{\Phi}\left(
\J^{ij}_{2,1,-1}X^{\mu}_{i}X^{\nu}_{j}+\frac{5}{3}\Delta^{\mu\nu}\J_{4,2,-1}
\right)\;.
\end{align}
The remaining dissipative forces in Eq.~(\ref{pimunu1}) ${\cal K}^{\mu\nu}$, ${\cal L}^{\mu\nu}$, etc. and transport coefficients $\delta^{\mu\nu}_{\Pi\Pi}$, $\delta^{\mu\nu\alpha\beta}_{\Pi\pi}$, etc. are given in Appendix C of Ref.~\cite{Bazow:2013ifa}. 

The viscous anisotropic hydrodynamic framework for a nonconformal system (with vanishing chemical potential) in a general (3+1)-dimensional framework is defined by Eqs.~(\ref{eq:be0}), (\ref{eq:par}), (\ref{eq:perp}), (\ref{2ndMomentFinal}), (\ref{Pi1}), and (\ref{pimunu1}). Structurally they reduce to Eqs.~(88) and (90) of Ref.~\cite{Bazow:2013ifa} when taking the limit $\Phi\to0$. The difference between the equations studied here and in \cite{Bazow:2013ifa} is that here we account for some of the bulk viscous effects non-perturbatively, by including them via the scalar field $\Phi$ already in the LO distribution function $f_\rs$. This leads to slight changes in the structure of the relaxation equations for 
$\Pit$ and $\pit^{\mu\nu}$ and also changes the values of the transport coefficients.

\section{(0+1)-dimensional expansion for a nonconformal system}
\label{sec:0+1d.}
\subsection{Reduced evolution equations}

In this section we present and solve the boost-invariant {\sc vaHydro} equations for a simplified situation without transverse expansion. In the following we will use the relaxation time approximation (RTA) for the scattering kernel,
\begin{equation}
\label{rta}
C[f]=-\frac{p\cdot u}{\tau_{eq}}\, \bigl[f({\bf p};\Lambda,\xi,\Phi){-}f_0(u{\cdot}p/T)\bigr] ,
\end{equation}
where $\tau_{eq}$ is the relaxation time, assumed to be momentum-independent. For transversely homogeneous systems undergoing boost-invariant longitudinal expansion, the Boltzmann equation (\ref{eq:be}) with an RTA collision kernel (\ref{rta}) can be solved exactly \cite{Florkowski:2013lya,Florkowski:2013lza,Baym:1984np}, and this can be used to determine the efficacy of various approximation schemes. In the situation just described there are no transverse derivatives, the comoving time derivative $\dot{A}=DA$ simply becomes $dA/d\tau$, and the shear stress tensor $\tilde\pi^{\mu\nu}$ is fully defined by a single non-vanishing component $\pit\equiv\pit^{z}_{z}=-\pit^{zz}$: at $z{\,=\,}0$, $\pit^{\mu}_{\ \nu}=\mathrm{diag}(0,-\pit/2,-\pit/2,\pit)$. For (0+1)-dimensional expansion with azimuthal symmetry we have the following simplifications:
\begin{eqnarray}
x_\lambda D_x u^\lambda&=&y_\lambda D_y u^\lambda=0 \;,\\
z_\lambda D_z u^\lambda&=&-\frac{1}{\tau}\;,\;\;\;\;\; \theta=\frac{1}{\tau}
\;.
\end{eqnarray}
Using this we can write the (0+1)-d viscous anisotropic hydrodynamic equations of motion as:
\begin{align}
\label{0thMoment_0p1}
\dot{\cal N}&=-\frac{{\cal N}}{\tau}-\frac{1}{\tau_{eq}}\left(
{\cal N}-{\cal N}_{eq}
\right)\;,
\\
\label{1stMoment_0p1}
\dot{\cal E}&=-\frac{1}{\tau}\left({\cal E}+{\cal P}_L+\Pit-\pit\right)\;,
\\
\label{2ndMoment_0p1}
\frac{d}{d\tau}\langle Ep^2_z\rangle-\frac{d}{d\tau}\langle Ep^2_x\rangle
&=\frac{1}{\tau}\left(\langle Ep^2_x\rangle-3\langle Ep^2_z\rangle\right)
+\frac{1}{\tau_{eq}}\left(\langle Ep^2_x\rangle-\langle Ep^2_z\rangle\right).
\end{align}
With some algebra, using the explicit functional form (\ref{isoN}) of the particle density (\ref{anisoN}) for Boltzmann statistics, the first two of these can be rewritten in terms of the parameters of the leading-order distribution $f_\rs$ as
\begin{align}
\label{xi_p_0+1d}
\frac{\dot\xi}{1{+}\Phi{+}\xi}-2\left(3{+}\frac{m}{\Lambda}\frac{K_1(m/\Lambda)}{K_2(m/\Lambda)} \right)\frac{\dot\Lambda}{\Lambda}
&+\left(\frac{2}{1{+}\Phi}{+}\frac{1}{1{+}\Phi{+}\xi}\right)\dot{\Phi} 
\nonumber \\
&=
\frac{2}{\tau}+2\Gamma\left(1-\frac{T}{\Lambda}\frac{K_2(m/T)}{K_2(m/\Lambda)}(1{+}\Phi)\sqrt{1{+}\Phi{+}\xi}\right)\;,
\\
\label{l_p_0+1d}
(\partial_\xi{\cal E}) \dot\xi + (\partial_\Lambda{\cal E}) \dot\Lambda 
+(\partial_\Phi{\cal E}) \dot\Phi 
&=
- \frac{1}{\tau}\left({\cal E} + {\cal P}_L+\Pit-\pit\right)\,,
\end{align}
where the partial derivatives of ${\cal E}$ on the left hand side of Eq.~(\ref{l_p_0+1d}) can be worked out from the explicit expression (\ref{anisoE}). The effective temperature $T$ in Eq.~(\ref{xi_p_0+1d}) is obtained from the dynamical Landau matching condition 
\begin{equation}
\label{dynLandau}
{\cal E}(\Lambda,\xi,\Phi;\,m)={\cal E}_{eq}(T;\,m) \,.
\end{equation}
Some additional work yields the evolution equations for $\Pit$ and $\pit$ in the form
\begin{align}
\dot{\Pit}&=-\Gamma\left(\frac{2{\cal P}_\perp{+}{\cal P}_L}{3}{-}{\cal P}_\mathrm{eq}{+}\Pit\right)
{+}\frac{m^2}{3\Lambda}\left(\J_{0,0,1}\frac{\dot\Lambda}{\Lambda} -\frac{1}{2}\J^{zz}_{0,0,-1}\dot{\xi}
-\frac{3}{2}\J_{2,1,-1}\dot{\Phi}
+\frac{1{+}\Phi{+}\xi}{\tau}\J^{zz}_{0,0,-1}\right) \nonumber\\
&-\lambda_{\Pi\Pi}\frac{\Pit}{\tau}-\lambda_{\Pi\pi}\frac{\pit}{\tau}\,,
\label{pi0p1}
\end{align}
\begin{align}
\dot{\pit}&=-\Gamma\left(\pit-\frac{2}{3}({\cal P}_L-{\cal P}_\perp)\right)
\nonumber\\
&+\frac{1}{\Lambda}\left[\left(\J^{zz}_{0,0,1}-\J_{2,1,1}\right)\frac{\dot\Lambda}{\Lambda}+\left(\frac{1{+}\Phi{+}\xi}{\tau}-\frac{\dot\xi}{2}\right)\left(\J^{zzzz}_{0,0,-1}{-}\J^{zz}_{2,1,-1}\right)
-\frac{3}{2}\left(\J^{zz}_{2,1,-1}{-}\frac{5}{3}\J_{4,2,-1}\right)\dot\Phi
\right] 
\nonumber\\
&+\lambda_{\pi\Pi}\frac{\Pit}{\tau}+\lambda_{\pi\pi}\frac{\pit}{\tau}\,.
\label{pizz0p1}
\end{align}

Equations (\ref{2ndMoment_0p1})-(\ref{pizz0p1}) form the coupled set of dynamical equations that must be solved for (0+1)-dimensional expansion. The $\J$ integrals appearing in the last two of these equations are defined in the Appendix. The terms $\langle Ep^2_x\rangle$ and $\langle Ep^2_z\rangle$ appearing in Eq.~(\ref{2ndMoment_0p1}) involve the transport coefficients $\psi^{ij}_\Pi$ and $\psi^{ijzz}_{\pi}$. From a formal point of view, it is nice to have analytic expressions for the transport coefficients. However, for nonconformal systems the ``shear-bulk'' coupling is rather complicated. For numerical purposes it is then easier to just use the parametrization of the non-equilibrium distribution function (\ref{q_expanded}) and numerically invert the matrix equation ${\cal A}\bm{b}=\bm{c}$ at each time step in the numerical integration to obtain the coefficients in the 14-moment approximation. We now show how to do this. For (0+1)-dimensional expansion with azimuthal symmetry we have the following simplifications:
\begin{eqnarray}
&&\bm{b}\equiv 
 \begin{pmatrix}
  \alpha & \beta & w & w_{11} & 0 & 0 & w_{22} & 0 & w_{33} 
 \end{pmatrix}^T \,,
\\ \label{81}
&&\bm{c}\equiv 
 \begin{pmatrix}
  0& 0 & \Pit & \pit/2 & 0 & 0 & \pit/2 & 0 & -\pit 
 \end{pmatrix}^T \,.
\end{eqnarray}
\begin{eqnarray}
x_\lambda D_x u^\lambda&=&y_\lambda D_y u^\lambda=0 \;,\\
z_\lambda D_z u^\lambda&=&-\frac{1}{\tau}\;,\;\;\;\;\; \theta=\frac{1}{\tau}
\end{eqnarray}
By defining the vector
\begin{equation}
\bm{\mathcal{J}}^{ii}\equiv 
 \begin{pmatrix}
  \J^{ii}_{1,0}& \J^{ii}_{2,0} & \J^{ii}_{3,0}+\J^{ii}_{3,1} 
  & \J^{iixx}_{1,0} & 0 & 0 & \J^{iixx}_{1,0} & 0 & \J^{iizz}_{1,0} 
 \end{pmatrix}^T \;,
\end{equation}
we can write
\begin{equation}
\langle Ep^2_i\rangle\equiv\I^{ii}_{1,0}+\vec{b}\cdot\vec{\cal J}^{ii}\;.
\end{equation}
Then taking the derivative 
\begin{equation}
\label{dEpi2dt}
\frac{d}{d\tau}\langle Ep^2_i\rangle=
\frac{d\I^{ii}_{1,0}}{d\tau}
-\left({\cal A}^{-1}\frac{d{\cal A}}{d\tau}\,\bm{b}\right){\cdot}\bm{\mathcal{J}}^{ii}
+\left({\cal A}^{-1}\bm{\hat{u}}\right){\cdot}\bm{\mathcal{J}}^{ii}\frac{d\Pit}{d\tau}
+\left({\cal A}^{-1}\bm{\hat{v}}\right){\cdot}\bm{\mathcal{J}}^{ii}\frac{d\pit}{d\tau}
+\vec{b}{\cdot}\frac{d\vec{\cal J}^{ii}}{d\tau}\;,
\end{equation}
where we introduced the vectors
\begin{eqnarray}
&&\bm{\hat{u}}\equiv 
 \begin{pmatrix}
  0 & 0 & 1 & 0 & 0 & 0 & 0 & 0 & 0 
 \end{pmatrix}^T \,,
\\ \label{81}
&&\bm{\hat{v}}\equiv 
 \begin{pmatrix}
  0& 0 & 0 & \frac{1}{2} & 0 & 0 & \frac{1}{2} & 0 & -1 
 \end{pmatrix}^T \,.
\end{eqnarray}
With the anisotropic form (\ref{eq:lo_df}) as the underlying LO distribution function, it is convenient to evolve the system in terms of the kinematical parameters $\xi$, $\Phi$, and $\Lambda$, rather than the macroscopic densities. Writing Eq.~(\ref{dEpi2dt}) in terms of $\dot\xi$, $\dot\Phi$, and $\dot\Lambda$
\begin{equation}
\frac{d}{d\tau}\langle Ep^2_i\rangle=
\psi^{ii}_{\xi}\dot{\xi}+\psi^{ii}_{\Phi}\dot{\Phi}
+\psi^{ii}_{\Lambda}\dot{\Lambda}
+\psi^{ii}_{\Pi}\dot{\Pit}+\psi^{ii}_{\pi}\dot{\pit}\;,
\end{equation}
where we have introduced the shorthand notation:
\begin{align}
\psi^{ii}_{a}&\equiv
\partial_{a}\I^{ii}_{1,0}
-\left({\cal A}^{-1}(\partial_{a}{\cal A})\,\bm{b}\right){\cdot}\bm{\mathcal{J}}^{ii}
+\bm{b} \cdot \partial_{a}\bm{\mathcal{J}}^{ii}\;,
\\
\psi^{ii}_{\Pi}&\equiv\left({\cal A}^{-1}\bm{\hat{u}}\right){\cdot}\bm{\mathcal{J}}^{ii}\;,
\\
\psi^{ii}_{\pi}&\equiv\left({\cal A}^{-1}\bm{v}\right){\cdot}\bm{\mathcal{J}}^{ii}
\;.
\end{align}
Then defining $\psi_{k}\equiv\psi^{zz}_{k}-\psi^{xx}_{k}$, where $k\in\{\xi,\Lambda,\Phi,\Pit,\pit\}$, and taking the $zz$-component of Eq.~(\ref{2ndMoment_0p1}), we get
\begin{equation}
\psi_{\xi}\dot{\xi}+\psi_{\Lambda}\dot{\Lambda}+\psi_{\Phi}\dot{\Phi}
+\psi_{\Pi}\dot{\Pit}+\psi_{\pi}\dot{\pit}
=\frac{1}{\tau}\left(\langle Ep^2_x\rangle-3\langle Ep^2_z\rangle\right)
+\frac{1}{\tau_{eq}}\left(\langle Ep^2_x\rangle-\langle Ep^2_z\rangle\right)\;.
\end{equation}
The transport coefficients appearing in Eqs.~(\ref{Pi1}) and (\ref{pimunu1}) can be written as
\begin{align}
\lambda_{\Pi\Pi}&=1+\frac{m^2}{3}\bm{b}_\Pi{\cdot}\bm{\mathcal{J}}_\Pi\;,\;\;\;\;\;
\lambda_{\Pi\pi}=\frac{m^2}{3}\bm{b}_\pi{\cdot}\bm{\mathcal{J}}_\Pi\;, 
\\
\lambda_{\pi\Pi}&=\frac{4}{3}+\bm{b}_\Pi{\cdot}\bm{\mathcal{J}}_\pi\;,\;\;\;\;\;
\lambda_{\pi\pi}=-\frac{7}{3}+\bm{b}_\pi{\cdot}\bm{\mathcal{J}}_\pi\;, 
\end{align}
where $\bm{b}_\Pi\equiv {\cal A}^{-1}\bm{\hat{u}}$, $\bm{b}_\pi\equiv {\cal A}^{-1}\bm{\hat{v}}$, and
\begin{eqnarray}
&&\tilde{\!\!\bm{\mathcal{J}}}_{\!\!\Pi}\equiv 
 \begin{pmatrix}
  \J^{zz}_{-2,0}, & \J^{zz}_{-1,0}, & \J^{zz}_{0,0}+\J^{zz}_{0,1}, & 
  \J^{xxzz}_{-2,0}, & 0, & 0, & \J^{xxzz}_{-2,0}, & 0, & \J^{zzzz}_{-2,0}  
 \end{pmatrix} \,,
\\ \label{JPi}
&&\tilde{\!\!\bm{\mathcal{J}}}_{\!\!\pi}\equiv 
 \left(
  \J^{zz}_{0,1}{-}\J^{zzzz}_{-2,0}, \J^{zz}_{1,1}{-}\J^{zzzz}_{-1,0},  
  \J^{zz}_{2,1}{+}\frac{5}{3}\J^{zz}_{2,2}{-}\J^{zzzz}_{0,0}{-}\J^{zzzz}_{0,1},
  \right.
  \nonumber\\
  &&\hspace{2cm}\left.
  \J^{xxzz}_{0,1}{-}\J^{xxzzzz}_{-2,0}, 0, 0, \J^{xxzz}_{0,1}{-}\J^{xxzzzz}_{-2,0}, 0, \J^{zzzz}_{0,1}{-}\J^{zzzzzz}_{-2,0}
 \right) \,.
\end{eqnarray}
%

\subsection{Numerical results}
\begin{figure}[t!]
\begin{center}
\includegraphics[width=\linewidth]{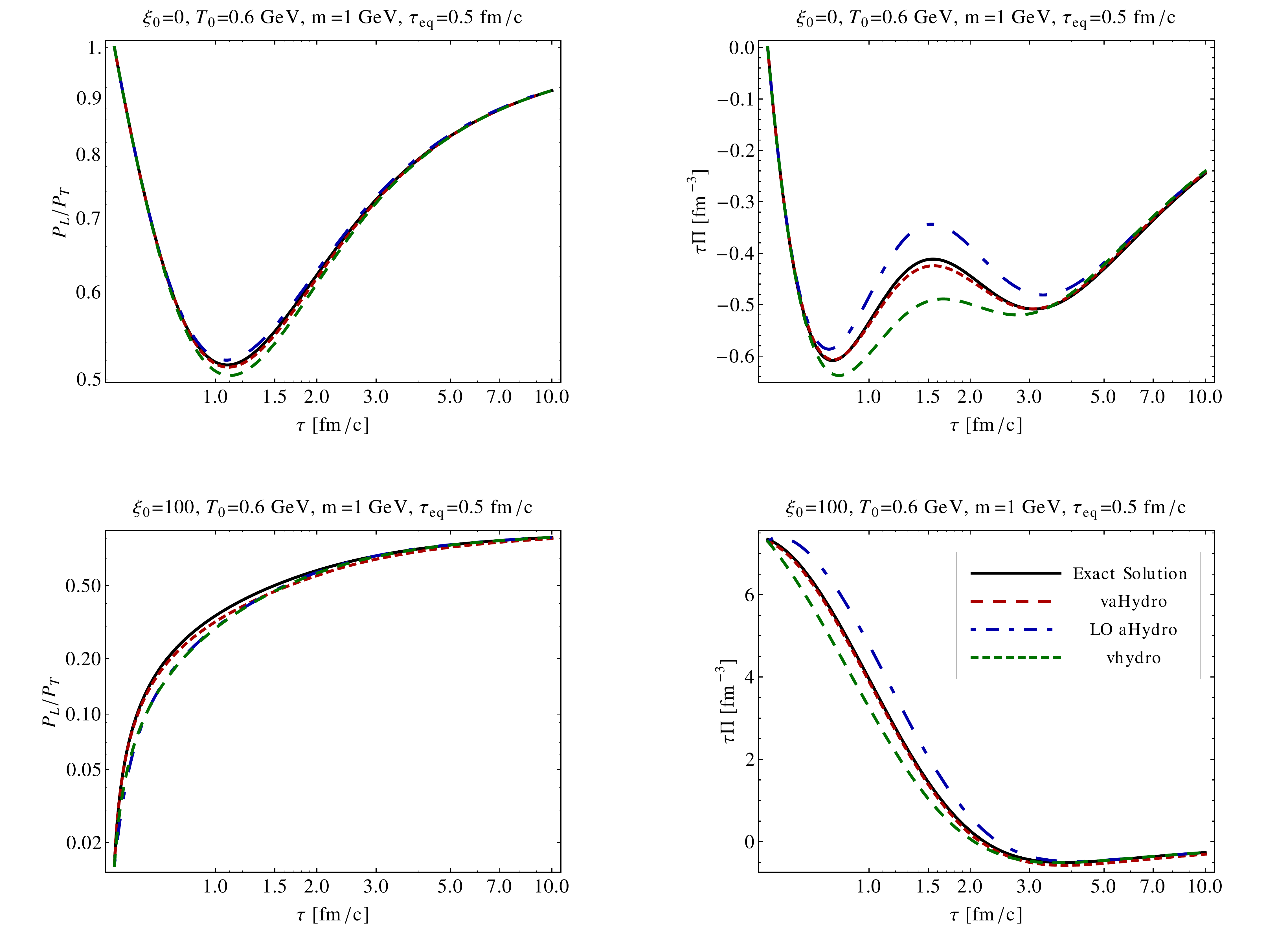}
\end{center}
\vspace{-7mm}
\caption{Ratio of the longitudinal to transverse pressure ${\cal P}_\perp/{\cal P}_L$ (left column) and the bulk viscous pressure $\Pi$ (right column). The top panels correspond to an initial anisotropy parameter $\xi_0=0$ whereas the bottom panels are for $\xi_0=100$. The black solid, red short-dashed, blue dashed-dotted, and green long-dashed lines are the results obtained from the exact solution of the Boltzmann equation, NLO anisotropic hydrodynamics ({\sc vaHydro}), LO anisotropic hydrodynamics ({\sc aHydro}), and second-order viscous hydrodynamics~\cite{Denicol:2010xn,Denicol:2012cn,Denicol:2012es}, respectively. The initial conditions in this figure are $T_0=600$\,MeV, $m=1$\,GeV, $\Pit_0=0$, $\pit_0=0$, $\tau_{eq}=0.5$\,fm/$c$, and $\tau_0=0.5$\,fm/$c$.}
\label{F1}
\end{figure}

\begin{figure}[t!]
\begin{center}
\includegraphics[width=\linewidth]{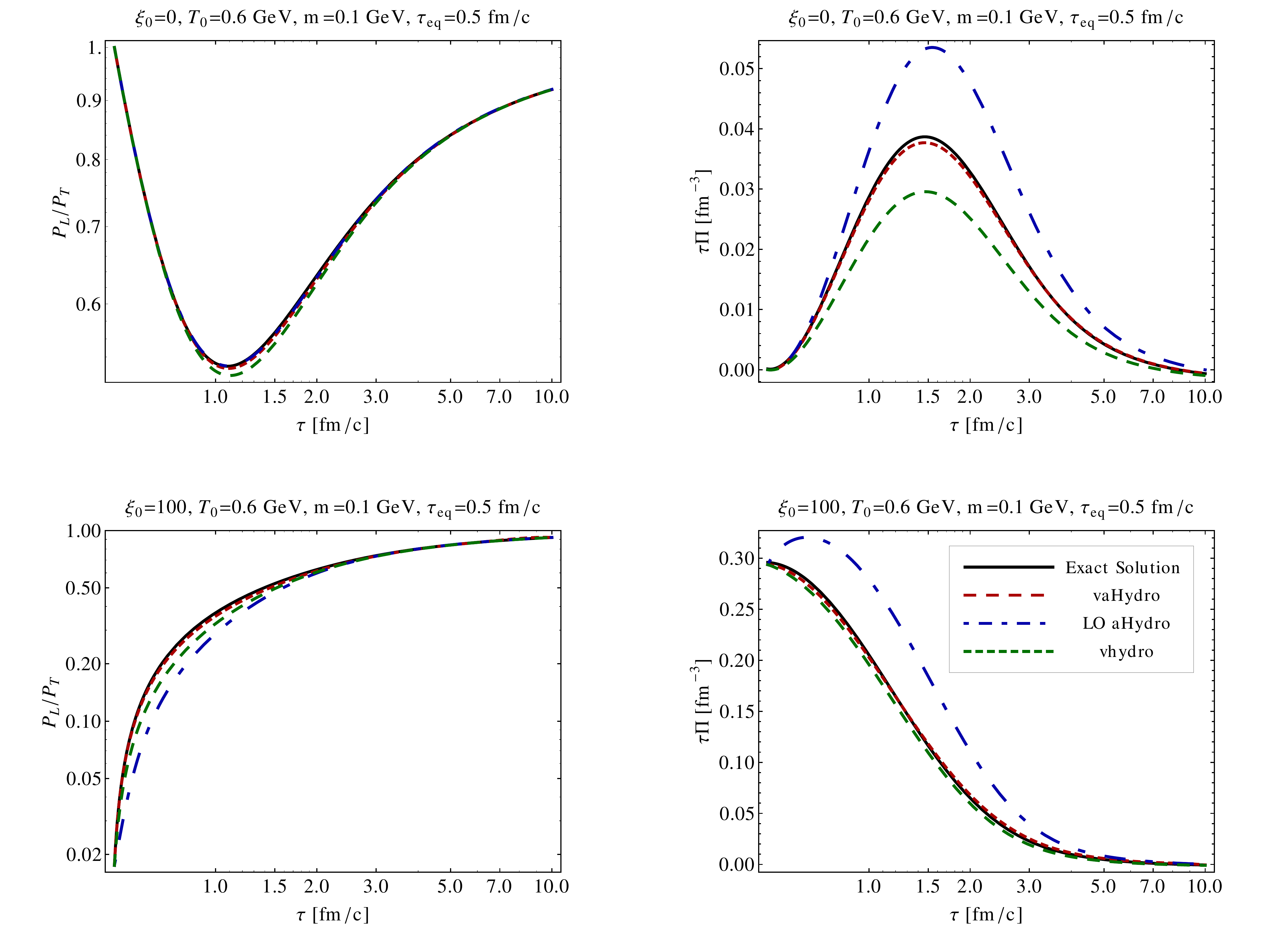}
\end{center}
\vspace{-7mm}
\caption{Similar to Fig.~\ref{F2}, but for a ten times smaller mass $m=0.1$\,GeV.}
\label{F2}
\end{figure}

\begin{figure}[t]
\begin{center}
\includegraphics[width=\linewidth]{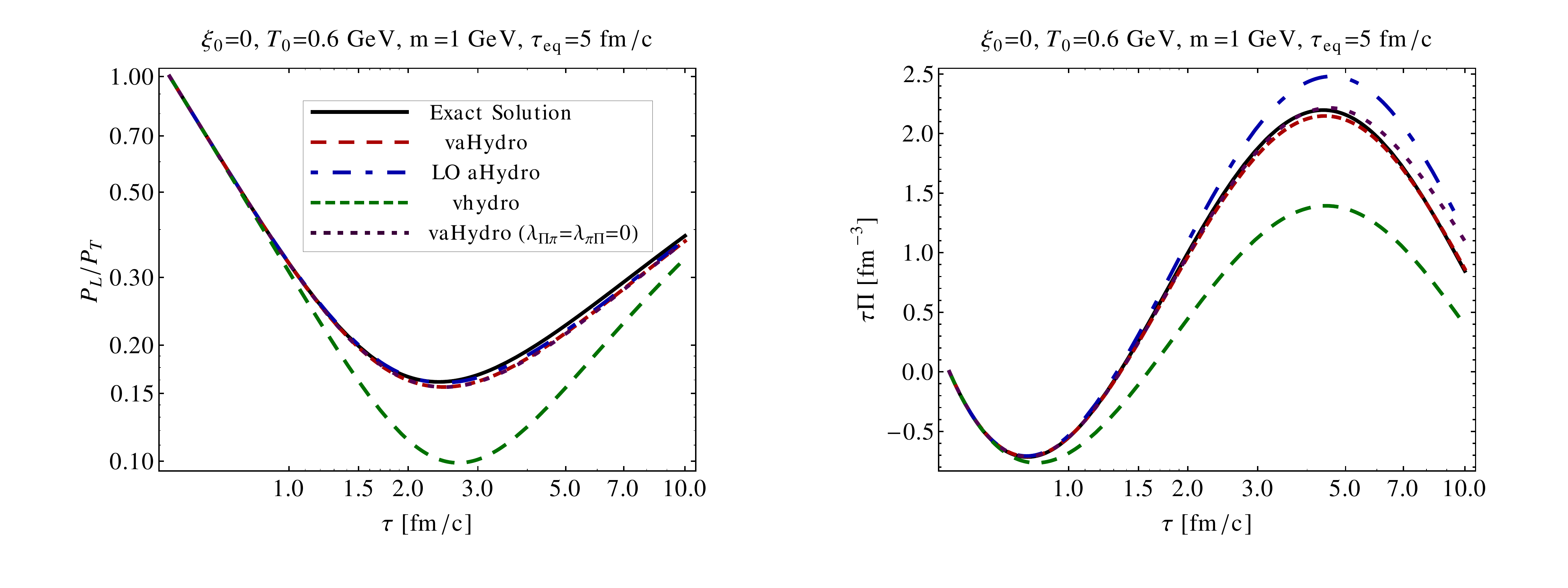}
\end{center}
\vspace{-7mm}
\caption{Similar to Fig.~\ref{F1}, but for $\tau_\mathrm{eq}{\,=\,}5$\,fm/$c$. An additional set of purple dotted curves shows the effect of setting the bulk-shear coupling terms in the evolution equations for $\Pit$ and $\pit$ to zero.}
\label{F3}
\end{figure}

In this subsection we solve for (0+1)-dimensional expansion the {\sc vaHydro} equations (\ref{2ndMoment_0p1})-(\ref{pizz0p1}) numerically and compare the resulting evolution histories for the macroscopic thermodynamic quantities with the corresponding moments of the related exact solution of the Boltzmann equation \cite{Florkowski:2014sfa}. We initialize the system at $\tau_0=0.5$\,fm/$c$ with $T_0=600$\,MeV, $\Pit_0=0$, and $\pit_0=0$.For simplicity and illustration, we assume a temperature independent relaxation time, exploring the cases  $\tau_{eq}=0.5$\,fm/$c$ in Figs.~\ref{F1} and \ref{F2} as well as the ten times larger value $\tau_{eq}=5$\,fm/$c$ in Fig.~\ref{F3}. 

In Figs.~\ref{F1}$-$\ref{F3} we plot in the left panels the evolution of the ratio ${\cal P}_\perp/{\cal P}_L$ between the longitudinal and transverse pressures, and in the right panels the bulk viscous pressure $\Pi$. For Figs.~\ref{F1} and \ref{F3} we assume particles of mass $m=1$\,GeV while in Fig.~\ref{F2} we use $m=0.1$\,GeV. The upper and lower panels in Figs.~\ref{F1} and \ref{F2} correspond to different initial momentum-space anisotropies: For the histories shown in the upper panels of Figs.~\ref{F1} and \ref{F2}, as well as those in Fig.~\ref{F3}, we assumed  initial momentum isotropy, $\xi_0=0$. The lower panels in Figs.~\ref{F1} and \ref{F2} start instead from a very anisotropic initial state with $\xi_0=100$. In all cases, the solid black line shows the results obtained from the exact solution of the Boltzmann equation. The short-dashed red lines represent our {\sc vaHydro} results while the dashed-dotted blue and long-dashed green curves show results from LO anisotropic hydrodynamics ({\sc aHydro}) \cite{Nopoush:2014pfa} and second-order viscous hydrodynamics ({\sc vHydro}) evaluated in the 14-moment approximation \cite{Denicol:2010xn,Denicol:2012cn,Denicol:2012es} for comparison.

We observe that in all cases the {\sc vaHydro} framework gives the best approximation to the exact solution for the longitudinal/transverse pressure ratio and the bulk viscous pressure. Especially during the early evolution stages, {\sc vaHydro} matches the exact solution almost perfectly while LO {\sc aHydro} and second-order viscous hydrodynamics \cite{Denicol:2010xn,Denicol:2012cn,Denicol:2012es} exhibit significant deviations. The improvement achieved by including in the dynamical description the additional dissipative flows generated by $\dft$ is particularly visible in the evolution of the bulk viscous pressure, caused by the non-vanishing particle mass. 

As discussed in \cite{Denicol:2014mca}, for massive particle systems coupling terms between the shear and bulk viscous pressures play an important role in the evolution of the viscous stress; in viscous hydrodynamics ({\sc vHydro}) these must be included explicitly at second order in an expansion around a locally isotropic momentum distribution \cite{Denicol:2010xn, Denicol:2012cn, Denicol:2012es} while anisotropic hydrodynamics ({\sc aHydro}), based on the ansatz (\ref{eq:lo_df}), captures their effects already at leading order, with similar precision. The dotted purple curves in Fig.~\ref{F3}, which were obtained by setting in Eqs.~(\ref{pi0p1}) and (\ref{pizz0p1}) the bulk-shear coupling coefficients $\lambda_{\Pi\pi}$ and $\lambda_{\pi\Pi}$ to zero, show that in our improved {\sc vaHydro} framework residual bulk-shear coupling terms between $\Pit$ and $\pit$ (due to $\dft$) play only a minor role, and only at late times. The main improvement over {\sc aHydro} and second-order {\sc vHydro} results from the other terms on the right hand sides of these equations, including the diagonal couplings $\lambda_{\Pi\Pi}$ and $\lambda_{\pi\pi}$.   

\section{Conclusions}
\label{sec:concl.}

In this paper we derived a generalization of the viscous anisotropic hydrodynamic framework \cite{Bazow:2013ifa} to systems with massive degrees of freedom, assuming a vanishing chemical potential. To test the efficacy of this extended {\sc vaHydro} formalism we applied it to a transversally homogeneous non-conformal system that undergoes boost-invariant longitudinal expansion (i.e. (0+1)-dimensional flow) for which there exists an exact solution of the RTA Boltzmann equation \cite{Florkowski:2014sfa, Florkowski:2014bba}. We tested its precision in this situation by comparing, over a wide range of particle masses, relaxation times and initial momentum anisotropy parameters, the numerical predictions of {\sc vaHydro} for the longitudinal/transverse pressure ratio and the bulk viscous pressure with the corresponding results obtained from the exact (0+1)-d solution of the RTA Boltzmann equation \cite{Florkowski:2014sfa,Florkowski:2014bba}, as well as with those from two other hydrodynamic expansion schemes, namely second-order viscous hydrodynamics in the 14-moment approximation ({\sc vHydro}) \cite{Denicol:2010xn,Denicol:2012cn,Denicol:2012es} and anisotropic hydrodynamics ({\sc aHydro}) \cite{Nopoush:2014pfa}. In all cases we found that {\sc vaHydro} agrees almost perfectly with the exact kinetic solution and presents a significant improvement over the other two hydrodynamic approaches. 


For massless theories, a powerful test for the efficiency of various hydrodynamic approximation schemes for the RTA Boltzmann equation is the amount of entropy generated by viscous heating during the evolution of the system \cite{Martinez:2012tu,Bazow:2013ifa}. Among all known hydrodynamic schemes, only {\sc aHydro} and {\sc vaHydro} reproduce this quantity qualitatively correctly in both the strong ($\tau_\mathrm{rel}\to0$) and weak ($\tau_\mathrm{rel}\to\infty$) coupling limits, and for (0+1)-d expansion {\sc vaHydro} does so almost perfectly. It would have been nice to perform this check also for the case of massive theories. However, for massive particles the entropy needs to be computed from kinetic theory (rather than using the particle density). We postpone this for future work.

\acknowledgments
We thank R. Ryblewski for providing us with his code for the numerical evaluation of the exact solution of the massive Boltzmann equation \cite{Florkowski:2014sfa}. This work was supported by the U.S. Department of Energy, Office of Science, Office of Nuclear Physics under Awards No. \rm{DE-SC0004286} and (within the framework of the JET Collaboration) \rm{DE-SC0004104}. 

\appendix
\section*{Appendix: Evaluation of the ``anisotropic" thermodynamic integrals}
\label{sec:jints}
In this Appendix we compute the auxiliary ``anisotropic" thermodynamic integrals $\I^{i_1\cdots i_k}_{nqr}$ and $\J^{i_1\cdots i_k}_{nqr}$ defined as
\begin{align}
\I^{i_1\cdots i_\ell j_1\cdots j_p}_{nqr} &\equiv
\frac{(-1)^q}{(2q+1)!!}\int dP\,E^{n-2q}
\left(\Delta^{\alpha\beta}p_\alpha p_\beta\right)^q E_\rs^r\,
p_{i_1}\cdots p_{i_\ell}p_{j_1}\cdots p_{j_p}f_\rs \label{Iint}\;,\\
\J^{i_1\cdots i_\ell j_1\cdots j_p}_{nqr} &\equiv
\frac{(-1)^q}{(2q+1)!!}\int dP\,E^{n-2q}
\left(\Delta^{\alpha\beta}p_\alpha p_\beta\right)^q E_\rs^r\,
p_{i_1}\cdots p_{i_\ell}p_{j_1}\cdots p_{j_p}f_\rs\tilde{f}_\rs\;,
\end{align}
where $i=1,2$ and $j=3$ denote the number of $p_i$ and $p_z$ components, respectively, and
$\tilde{f}_\rs{\,=\,}1{-}af_\rs$. To work out the derivative of $\I^{i_1\cdots i_\ell j_1\cdots j_p}_{nqr}$ with respect to $\xi$, $\Phi$ and $\Lambda$ we use the following relations:
\begin{align}
\partial_\xi\left(E^r_\rs f_\rs\right)&=\frac{p_z^2}{2E_\rs}\left(\frac{r}{E_\rs}-\frac{1}{\Lambda}\right)f_\rs\tilde{f}_\rs\;,
\\
\partial_\Phi\left(E^r_\rs f_\rs\right)&=-\frac{\Delta^{\alpha\beta}p_\alpha p_\beta}{2E_\rs}\left(\frac{r}{E_\rs}-\frac{1}{\Lambda}\right)f_\rs\tilde{f}_\rs\;,
\\
\partial_\Lambda f_\rs&=\frac{E_\rs}{\Lambda^2}f_\rs\;.
\end{align}
This allows us to write
\begin{align}
\partial_\xi\I^{i_1\cdots i_\ell j_1\cdots j_p}_{nqr}=
\frac{1}{2}\left(r\J^{i_1\cdots i_\ell j_1\cdots j_{p+2}}_{n,q,r-2}-\frac{1}{\Lambda}\J^{i_1\cdots i_\ell j_1\cdots j_{p+2}}_{n,q,r-1}\right)
\end{align}
\begin{align}
\partial_\Phi\I^{i_1\cdots i_\ell j_1\cdots j_p}_{nqr}=
\frac{1}{2}\left(r\J^{i_1\cdots i_\ell j_1\cdots j_{p}}_{n+2,q,r-2}-\frac{1}{\Lambda}\J^{i_1\cdots i_\ell j_1\cdots j_{p}}_{n+2,q,r-1}\right)
\end{align}
\begin{align}
\partial_\Lambda\I^{i_1\cdots i_\ell j_1\cdots j_p}_{nqr}=
\frac{1}{\Lambda^2}\J^{i_1\cdots i_\ell j_1\cdots j_p}_{n,q,r+1}
\end{align}
We parametrize the four-momenta in hyperbolic coordinates,
\begin{equation}
p^\mu=(m_\perp\cosh y,p_\perp\cos\phi,p_\perp\sin\phi,m_\perp\sinh y)\;,
\end{equation}
where $m_\perp^2\equiv m^2+p_\perp^2$ is the transverse mass, with integration measure
\begin{equation}
dP=\frac{dy\, m_\perp dm_\perp d\phi}{(2\pi)^3}.
\end{equation}
Then 
\begin{align}
\I^{i_1\cdots i_\ell j_1\cdots j_p}_{nqr} 
&\equiv \frac{(-1)^q}{(2q+1)!!} \frac{1}{(2\pi)^3}
\int dy\, m_\perp dm_\perp d\phi \,(m_\perp\cosh y)^{n-2q}
\left(m^2{-}m_\perp^2\cosh^2 y\right)^q
\nonumber\\
&\times
\left[m_\perp^2\left((1{+}\Phi)\cosh^2y+\xi\sinh^{2}y\right)-m^2\Phi \right]^{r/2} 
\left(m_\perp^2{-}m^2\right)^{\ell/2}\bar{p}_{i_1}\cdots\bar{p}_{i_\ell}
(m_\perp\sinh{y})^p
\nonumber\\
&\times 
f_0\left(\frac{1}{\Lambda}
\sqrt{m_\perp^2\left((1{+}\Phi)\cosh^2y+\xi\sinh^{2}y\right)-m^2\Phi}
\right),
\end{align}
where we defined the scaled transverse Cartesian momentum components $\bar{p}_i\equiv p_i/p_\perp$, such that $\bar{p}_x=\cos\phi$ and $\bar{p}_y=\sin\phi$. We now define dimensionless parameters $\hat{m}_\perp\equiv m_\perp/\Lambda$ and $\hat{m}\equiv m/\Lambda$ which results in 
\begin{equation}
\begin{split}
\I^{i_1\cdots i_\ell j_1\cdots j_p}_{nqr} \equiv
\frac{(-1)^q}{(2q+1)!!}\frac{\Lambda^{n+\ell+p+r+2}}{2\pi^2}\,
\Phi^{i_1\cdots i_\ell}\int dy\,\cosh^{n-2q}y\sinh^{p}y\,
{\cal H}_{nq\ell p r}\left(y,\xi,\Phi;\hat{m}\right),
\end{split}
\end{equation}
where
\begin{align}
&\Phi^{i_1\cdots i_\ell}\equiv\int_0^{2\pi}\frac{d\phi}{2\pi}\bar{p}_{i_1}\cdots\bar{p}_{i_\ell} 
\\
&{\cal H}_{nq\ell p r}\left(y,\xi,\Phi;\hat{m}\right)
\equiv\int_{\hat{m}}^\infty d\hat{m}_\perp \hat{m}_\perp^{n-2q+p+1}
\left[\hat{m}_\perp^2\left((1{+}\Phi)\cosh^2y+\xi\sinh^{2}y\right)-\hat{m}^2\Phi \right]^{r/2}
\\\nonumber
&\qquad\times\left(\hat{m}^2{-}\hat{m}_\perp^2\cosh^2 y\right)^q
\left(\hat{m}_\perp^2{-}\hat{m}^2\right)^{\ell/2}\,
f_0\left(\sqrt{\hat{m}_\perp^2\left((1{+}\Phi)\cosh^2y+\xi\sinh^{2}y\right)-\hat{m}^2\Phi}\right).
\end{align}
The same decomposition follows for $\J^{i_1\cdots i_\ell j_1\cdots j_p}_{nqr}$:
\begin{equation}
\begin{split}
\J^{i_1\cdots i_\ell j_1\cdots j_p}_{nqr} \equiv
\frac{(-1)^q}{(2q+1)!!}\frac{\Lambda^{n+\ell+p+r+2}}{2\pi^2}\,
\Phi^{i_1\cdots i_\ell}\int dy\,\cosh^{n-2q}y\sinh^{p}y\,
\tilde{\cal H}_{nq\ell p r}\left(y,\xi,\Phi;\hat{m}\right),
\end{split}
\end{equation}
where $\tilde{\cal H}_{nq\ell p r}$ takes into account quantum statistics and is obtained by making the substitution $f_0(\cdot)\to f_0(\cdot)\tilde{f}_0(\cdot)$. We note that in the classical limit ($a{\,=\,}0$) the two functions are identical, $\I^{i_1\cdots i_\ell j_1\cdots j_p}_{nqr}(a{=}0)=\J^{i_1\cdots i_\ell j_1\cdots j_p}_{nqr}(a{=}0)$.


\bibliography{nonconformal_vahydro}

\end{document}